\documentclass[12pt]{iopart}

\usepackage{iopams}
\usepackage{graphicx}
\usepackage{amsfonts}
\usepackage{amsbsy}
\usepackage{amssymb}
\usepackage{color}

\begin{document}

\title{Einstein relation in superdiffusive systems}
 
\author{G Gradenigo, A Sarracino} 
\address{Istituto dei Sistemi Complessi - CNR and Dipartimento di Fisica, Universit\`a Sapienza 
- p.le A. Moro 2, 00185, Roma, Italy} 
\author{D Villamaina}
\address{Universit\'e Paris Sud - CNRS - LPTMS, UMR 8626, Orsay 91405, France} 
\author{A Vulpiani} 
\address{Dipartimento di Fisica, Universit\`a Sapienza and Istituto dei Sistemi Complessi - CNR 
- p.le A. Moro 2, 00185, Roma, Italy}

\ead{ggradenigo@gmail.com, alessandro.sarracino@roma1.infn.it\\
dario.villamaina@lptms.u-psud.fr, angelo.vulpiani@roma1.infn.it}


\begin{abstract}
We study the Einstein relation between diffusion and response to
an external field in systems showing superdiffusion. In particular, we
investigate a continuous time L\'evy walk where the velocity remains
constant for a time $\tau$ with distribution $P_\tau(\tau) \sim
\tau^{-g}$. At varying $g$ the diffusion can be standard or anomalous;
in spite of this, if in the unperturbed system a current is absent,
the Einstein relation holds. In the case where a current is present
the scenario is more complicated and the usual Einstein relation
fails. This suggests that the main ingredient for the breaking of the
Einstein relation is not the anomalous diffusion but the presence of a
mean drift (current).
\end{abstract}

\maketitle 

\section{Introduction}

The fluctuation-dissipation relation (FDR) is one of the fundamental
results of the statistical physics~\cite{K66}.  In his celebrated work
on the Brownian motion, Einstein gave the first example of FDR.  In
the absence of external forcing, at large times, one has
\begin{equation}
\label{1}
\langle x(t) \rangle=0 \qquad \textrm{and} \qquad
\langle x^2(t) \rangle \simeq 2 D t,
\end{equation}
where $x$ is the position of the colloidal particle, $D$ is the
diffusion coefficient and the average refers to the unperturbed
dynamics.  If a small constant external force $\mathcal{E}$ is applied, one
obtains a linear drift
\begin{equation}
\label{2}
\langle x(t) \rangle_\mathcal{E} 
 \simeq \mu\mathcal{E}  t,
\end{equation}
where $\langle\ldots\rangle_\mathcal{E}$ is the average on the
perturbed system, and $\mu$ is the mobility.  The remarkable result
obtained by Einstein is the proportionality between $\langle x(t)
\rangle_\mathcal{E}$ and the mean square displacement (MSD) $\langle
x^2(t) \rangle$:

\begin{equation}
\label{3}
{\langle x^2(t) \rangle  \over \langle x(t) \rangle_\mathcal{E}}=\frac{2}{\beta \mathcal{E}},
\end{equation}
namely $\mu = \beta D$, where $\beta$ is the inverse temperature and
we have taken the Boltzmann constant $k_B=1$.

In the last decades many researches have been devoted to the study of
anomalous diffusion, where, instead of~(\ref{1}), one has
\begin{equation}
\label{4}
\langle x^2(t)\rangle \sim t^{2 \nu} \qquad \mbox{with} \qquad  \nu \ne
1/2,
\end{equation}
see for instance~\cite{BG90,GSGWS96,CMMV99,MK00,BC05,KS11}.  The case
$\nu<1/2$ is called subdiffusion while if $\nu>1/2$ we are in the
presence of superdiffusion.  It is quite natural to wonder about the
validity of the FDR in these anomalous situations. Important results showing
the proportionality between $\langle x(t) \rangle_\mathcal{E}$ and
$\langle x^2(t) \rangle$, which we refer to as the Einstein relation
``at equilibrium'', have already been obtained for both
superdiffusive~\cite{BF98,JMF99,KS11} and
subdiffusive~\cite{BG90,MBK99,KS11} anomalous dynamics.  Differently,
the situation where the drift is compared to the mean square
displacement in a state which is already ``out of equilibrium'',
either due to a current~\cite{VSGPV11} or due to
dissipation~\cite{VPV08}, has been studied only in the subdiffusive
cases.

The aim of this letter is to discuss the validity of the Einstein
relation in equilibrium and out-of-equilibrium situations in the
presence of superdiffusive dynamics.  In particular, we consider a
L\'evy walk collision process~\cite{BKLNP98} and we show that the
Einstein relation is violated when the perturbation is applied on a
reference state where a current is already present.

\section{The model}\label{sec:model}

We consider an ensemble of probe particles of mass $m$ endowed with
scalar velocity $v$ and position $x$. Each probe particle only
interacts with particles of mass $M$ and velocity $V$ extracted from
an equilibrium bath at temperature $T$. We assume that the scattering
probability does not depend on the relative velocity between the probe
particle and the colliders, as, for instance, in the case of
Maxwell-molecule models~\cite{E81}. Velocity of the probe particle
changes from $v$ to $v'$ at each collision, according to the rule:
\begin{equation} 
v'= \gamma v + (1- \gamma) V, 
\label{collrule}
\end{equation}
where $\gamma=(\zeta-\alpha)/(1+\zeta)$, with $\zeta=m/M$, and
$\alpha$ is the coefficient of restitution ($\alpha=1$ for an elastic
collision). The velocity $V$ of the bath particles is a random
variable generated from a Gaussian distribution with zero mean and
variance $T/M$:
\begin{equation} 
P_S(V)=\sqrt{\frac{M}{2 \pi T }} \exp\left\{-\frac{M}{2 T} V^2\right\}. \\
\label{scatt}
\end{equation}
The elementary step of the dynamics is made by: i) a flight,
$x(t+\tau)=x(t)+v'\tau+1/2\mathcal{E}\tau^2$, where $x(t)$ is the
position of the probe particle at time $t$, with $\tau$ taken from a
distribution $P_\tau(\tau)$ and $\mathcal{E}$ a constant
acceleration, followed by ii) a collision $v'=\gamma v + (1- \gamma)
V$, with $V$ taken from the Gaussian distribution~(\ref{scatt}).  In
the specific case $\alpha=1$ and $M=m$, one has $\gamma=0$ and the
collision rule~(\ref{collrule}) results in a random update of the
velocity according to the distribution~(\ref{scatt}). The duration of
each flight, $\tau$, is an independent identically distributed random
variable with probability

\begin{equation}
P_\tau(\tau)\sim\tau^{-g}, 
\label{levy}
\end{equation}
with $g>1$. This kind of process is called L\'evy walk collision
process~\cite{BKLNP98}, and \emph{may} be interpreted as due to
scattering centers randomly distributed on a fractal spatial
structure, as for instance in the case of molecular diffusion in porous
media~\cite{L97}.  If $\alpha \neq 1$ or $m \neq M$, with $\alpha\ne
\zeta$, a dependence on the last velocity before the collision
remains. In this case velocity correlations can be measured in the
system, as discussed in Section~\ref{velcor}.  

According to the dynamic rules of the process described above the
displacement of the probe particle is always finite in a finite
time. The anomalous dynamics of such a model has been studied
in~\cite{ACMV00}, showing that the process is an example of ``strong''
anomalous diffusion, namely that it is not possible to find a scaling
for the probability density function (PDF) of displacements. Such a
collision process becomes a standard diffusive system when $P_\tau$
decays fast enough: in this regime the dynamics is qualitatively
equivalent to that of a system with exponential $P_\tau$, studied for
instance in~\cite{AP10,GPSM12,BT12}.  L\'evy walk collision processes
have been thoroughly studied, see for instance~\cite{KS11}, where the
behavior of the mean square displacement has been obtained
analytically.

We recall here a simple argument already presented in~\cite{ACMV00} to
study the asymptotic behavior of higher order moments of displacement,
and that can be easily applied also to the case with an external
perturbing field that we are going to discuss here in
Section~\ref{pert}.  In order to obtain in a simple way the dominant
asymptotic behavior of $\langle x^2(t)\rangle$, we introduce a cut-off
$t_c$ for $P_\tau(\tau)$:

\begin{equation}\label{cutoff}
P_\tau(\tau) \sim \left\{ \begin{array}{ll}
\tau^{-g} & \textrm{if $\tau<t_c$}\\
0 & \textrm{if $\tau>t_c$.}\\ 
\end{array} \right.
\end{equation}
Assuming $x=0$ as initial condition for each trajectory, the mean
square displacement after the time $t$, where $N(t)$ collisions
occurred, can be written in full generality as:
\begin{equation}
  \langle x^2(t) \rangle = \left\langle \left[ \sum_{i=1}^{N(t)} v_i
    \tau_i \right]^2 \right\rangle \simeq  \sum_{i=1}^{\overline{N}(t)}\left\langle
  v^2_i \tau^2_i \right\rangle + 2\overline{N}(t)  \sum_{i=1}^{\overline{N}(t)}\left\langle
  v_iv_0 \tau_i\tau_0 \right\rangle.
  \label{MSD0}
\end{equation}
Here, $v_i$ denotes the velocity of the probe particle after the
$i$-th collision, $\tau_i$ is the time elapsed between the collisions $i$
and $i+1$ and $\overline{N}(t)$ is the average number of collisions
occurred up to time $t$.  The average $\langle\cdots\rangle$ is taken
over the distributions~(\ref{scatt}) and~(\ref{levy}). From
Eq.~(\ref{cutoff}) we have, for $n+1-g>0$,
\begin{equation}
\langle \tau^{n} \rangle_c \sim t_{c}^{n+1-g},
\label{taun}
\end{equation} 
where $\langle\cdots\rangle_c$ denotes an average over
the distribution~(\ref{cutoff}) with the cut-off $t_{c}$.

We start by considering the case with independent velocities $v_i$,
corresponding to a choice of parameters such that $\gamma=0$. Then,
estimating $\langle x^2(t) \rangle$ at a time $t\gg t_c$, so that the
average number of collisions along the trajectory can be approximated
to $\overline{N}(t)\simeq t/\langle \tau \rangle_c$, and considering
that the cross terms in Eq.~(\ref{MSD0}) are zero, we can write

\begin{equation}
  \langle x^2(t) \rangle \simeq \frac{t}{\langle \tau \rangle_c} \langle v^2\rangle \langle \tau^2 \rangle_c.
  \label{MSD1}
\end{equation}
In the case $g>3$, both $\langle \tau \rangle_c$ and $\langle \tau^{2}
\rangle_c$ are finite, even in the limit $t_c\rightarrow \infty$, so
that we find the simple diffusive behavior $\langle x^2(t) \rangle
\sim t$.  For $t\lesssim t_c$ and $1<g<3$, instead of~(\ref{MSD1}) we
expect

\begin{equation}
 \langle x^2(t) \rangle \sim t^{2\nu}.
\label{2nu}
\end{equation}
One can easily find the exponent $\nu$ with a matching
argument. Comparing~(\ref{MSD1}) and~(\ref{2nu}) at $t\sim t_c$ and
using~(\ref{taun}), we obtain $\nu=1/2$ for $g>3$, $\nu=2-g/2$ for
$2<g<3$, and $\nu=1$ for $1<g<2$ (logarithmic corrections appear at
the values $g=2$ and $g=3$~\cite{KS11}).

\begin{figure}[t!]
\center
\includegraphics[width=.7\columnwidth,clip=true]{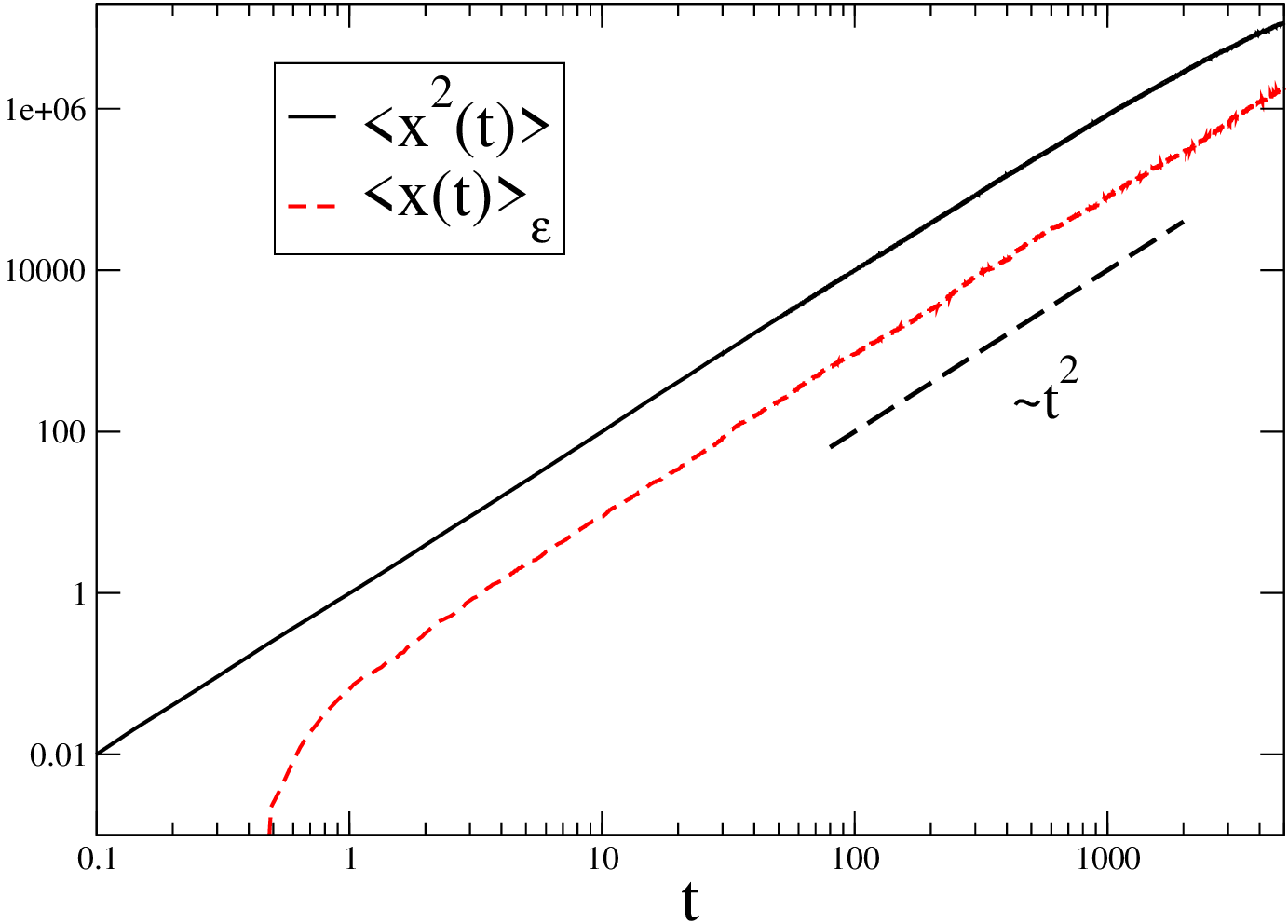}
\caption{Log-Log plot of the MSD in the case of
  superdiffusive dynamics with $g=3/2$, $\langle x^2(t)\rangle$ (black
  line), and of the drift $\langle x(t)\rangle_{\mathcal{E}}$ (red
  dashed line) in the case of uniform external field
  $\mathcal{E}=1.5$. Both quantities behave as $\sim t^2$ and the
  Einstein relation is verified.}
\label{fig0}
\end{figure}

\section{Einstein relation}
\label{pert}

 The main concern of our study is the question about the validity of
 the Einstein relation for superdiffusive anomalous dynamics.  In
 particular such a relation can be checked in two different
 situations:

\begin{description}
\item[(A)] The drift due to the external force is compared with the
  MSD of the probe particle in the absence of any pulling force.  This
  case corresponds to a fluctuation-dissipation experiment realized by
  switching on from zero the external perturbation.

\item[(B)] The drift is compared to the MSD in a state where a current
  is already present. This procedure corresponds to increase the
  intensity of the perturbation in a state already perturbed and
  compare the average current in this state with the fluctuations in
  the initial reference state.
\end{description}

In the following, we will refer to situation (A) as a test of the
fluctuation-dissipation relation at \emph{equilibrium} while to case
(B) as a test \emph{out of equilibrium}. We will show that these two
cases are very different.

\subsection{Perturbation of a state without current} 
\label{velcor}

The argument used in Sec.~\ref{sec:model} to study the MSD can be
applied to the drift, yielding

\begin{equation}
\hspace{-0.5cm}\langle x(t) \rangle_{\mathcal{E}} = \langle \sum_{i=1}^{N(t)} \left(
v_i \tau_i + \frac{\mathcal{E}}{2} \tau_i^2 \right)\rangle
=\frac{t}{\langle \tau \rangle_{c}} \left[ \langle \tau \rangle_{c}
  \langle v \rangle + \frac{\mathcal{E}}{2} \langle \tau^2
  \rangle_{c}\right] =\frac{\mathcal{E}}{2} \frac{t}{\langle \tau
  \rangle_{c}} \langle \tau^2 \rangle_{c},
\label{DRIFT0}
\end{equation}
which perfectly matches the result for the MSD found in
Eq.~(\ref{MSD1}).  Therefore, when we perturb an equilibrium state,
namely a state without currents, we have for any value of $g>1$
\begin{equation}
\frac{\langle x^2(t) \rangle}{\langle x(t) \rangle_{\mathcal{E}}}
= \textrm{const}.
\label{einstein}
\end{equation}
Let us note that the Einstein relation holds quite generally, namely
it persists even if we make our process non-Markovian by allowing some
memory across collisions, that is, we put $\gamma>0$ in the collision
rule Eq.~(\ref{collrule}), preventing a complete reshuffling of
velocities. In this case the MSD reads as:
\begin{equation}
\langle x^2(t) \rangle = \sum_{i,j}^{\overline{N}(t)} \langle v_i \tau_i v_j
\tau_j \rangle \sim \frac{t}{\langle \tau \rangle_c} \left[ \langle
  v^2\rangle \langle \tau^2 \rangle_c + 2\langle \tau\rangle_c^2\sum_i^{\overline{N}(t)} \langle v_i v_0
  \rangle  \right],
\label{corr}
\end{equation}
because the collision times are still not dependent on the velocity
but correlations between the velocities must be taken into account.
Note that, even in the presence of non-independent $\{v_i\}$, if
$\sum_i^n \langle v_i v_0 \rangle $ yields a finite contribution at
large times the second term on the right of Eq.~(\ref{corr}) grows as
$\langle\tau\rangle_c^2$, namely it is subdominant compared to
$\langle \tau^2\rangle_c$ in all situations.  If the exponent of the
flight time distribution is $1<g<2$, then $\langle \tau^2\rangle_c\sim
t^{3-g} > t^{4-2g} \sim \langle\tau\rangle_c^2$; if $2<g<3$ we have
that $\langle\tau\rangle_c^2$ is a finite number, while for $g>3$ we
recover a simple diffusive behavior for both the MSD and the drift.
In particular, in the case $g>3$ the presence of correlations of
velocities amounts to a renormalization of the diffusion coefficient.

We conclude that the asymptotic behavior of the MSD is the same 
as in Eq.~(\ref{MSD1}).  From Eq.~(\ref{DRIFT0}) one can see that no cross
correlations between velocities at different times appear, so that the
drift is not influenced by velocities correlations across
collisions. Therefore, we can conclude that the Einstein relation is
preserved for all values of the exponent $g$ of the flight times
distribution.  This example shows how the Einstein relation is a
stable result, valid under quite general conditions also in the presence
of anomalous dynamics~\cite{KS11,BPRV08,VSGPV11}.

\subsection{Perturbation of a state with a current}

\begin{figure}[t!]
\center
\includegraphics[width=.7\columnwidth,clip=true]{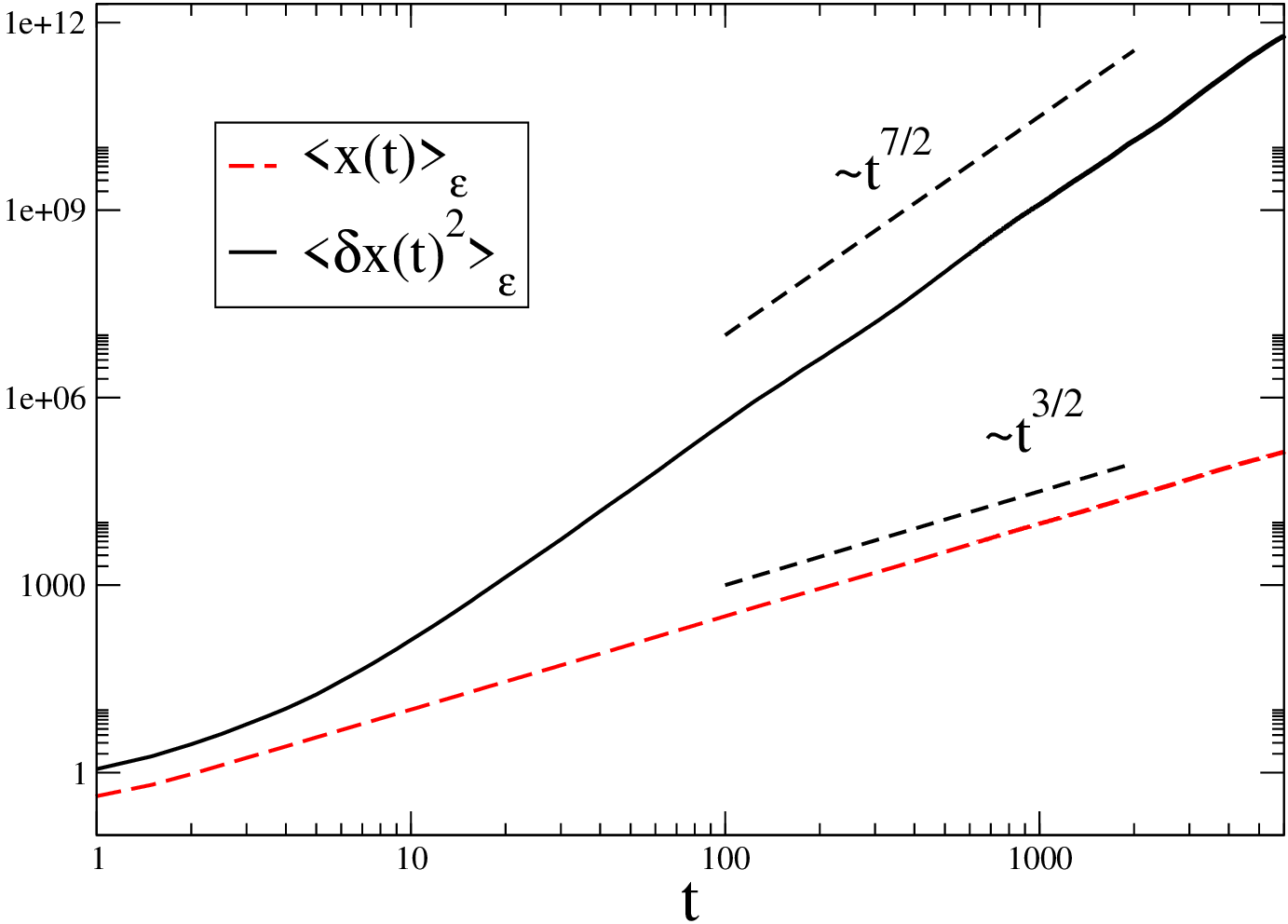}
\caption{Log-Log plot of the MSD (black line) in
  the presence of a constant external field $\mathcal{E}=1.5$ and
  drift (red dashed line) with the same value of the field, in the
  case of superdiffusive dynamics with $g=5/2$. We observe a breaking
  of the Einstein relation with leading asymptotic behaviors,
  $\langle [\delta x(t)]^2 \rangle_{\mathcal{E}} \sim t^{7/2}$, see
  Eq.~(\ref{PERTURBEDmsd1}), and $\langle x(t) \rangle_{\mathcal{E}}
  \sim t^{3/2}$, see Eq.~(\ref{DRIFT0}).}
\label{fig1}
\end{figure}

In order to study the Einstein relation out of equilibrium, let us
first consider a simple Gaussian process, namely the Brownian motion
of a colloidal particle when we add a constant force pulling the
particle. In this case it is sufficient to replace the MSD around the
average position $\langle [\delta x(t)]^2 \rangle_{\mathcal{E}} =
\langle x^2(t)\rangle_{\mathcal{E}}-\langle
x(t)\rangle^2_{\mathcal{E}}$ for the simple MSD, in order to recover
the Einstein relation with the drift $\langle x(t)
\rangle_{\mathcal{E}+\delta\mathcal{E}}-\langle x(t)
\rangle_{\mathcal{E}}$ (here
$\langle\cdots\rangle_{\mathcal{E}+\delta\mathcal{E}}$ denotes the
average over a state where the further perturbation
$\delta\mathcal{E}$ is applied).  In what follows we consider that a
non trivial violation of the Einstein relation happens when also the
MSD around the drift lacks the proportionality with the drift itself.
This is indeed the case of superdiffusive dynamics.  

For simplicity we will refer to the case $\gamma=0$ but the physical
picture remains the same also when the case with memory is
considered. In our model, by applying the constant field
$\mathcal{E}>0$, we have:
\begin{eqnarray}
\langle x^2(t) \rangle_{\mathcal{E}} &=& \langle \left[
  \sum_{i=1}^{N(t)} \left( v_i \tau_i + \frac{\mathcal{E}}{2} \tau_i^2 \right)
  \right]^2 \rangle_{t_c}  \nonumber \\
&\simeq& \frac{\mathcal{E}^2}{4}
t^2 \frac{\langle \tau^2 \rangle_c^2}{\langle \tau
    \rangle_c^2}+ t \left( \frac{\mathcal{E}^2}{4}\frac{\langle \tau^4 \rangle_c - \langle
    \tau^2 \rangle^2_c}{\langle \tau \rangle_c} + \frac{\langle v^2 \rangle \langle \tau^2
    \rangle_c}{\langle \tau \rangle_c}\right)  \\
\langle [\delta x(t)]^2 \rangle_{\mathcal{E}} &=& \langle
x^2(t)\rangle_{\mathcal{E}}-\langle x(t)\rangle^2_{\mathcal{E}} \simeq t
\left( \frac{\mathcal{E}^2}{4} \frac{\langle \tau^4 \rangle_c - \langle \tau^2 \rangle^2_c}{\langle \tau \rangle_c} +
 \frac{\langle v^2 \rangle \langle \tau^2 \rangle_c}{\langle \tau
  \rangle_c}\right).
\label{PERTURBEDmsd0}
\end{eqnarray}
In the case $2<g<3$, namely when the the distribution $P_\tau(\tau)$
has finite mean and infinite variance, the diffusion around the
average position behaves asymptotically as
\begin{equation}
\langle [\delta x(t)]^2 \rangle_{\mathcal{E}} \simeq t \left(
  \frac{\mathcal{E}^2}{4}\frac{t_c^{5-g}-t_c^{6-2g}}{\langle \tau \rangle_c}+
  \frac{\langle v^2\rangle t_c^{3-g}}{\langle  \tau \rangle_c} \right). 
\label{PERTURBEDmsd1}
\end{equation}
Considering for instance the case $g=5/2$, by applying the matching
argument to Eq.~(\ref{PERTURBEDmsd1}), we find that the leading
behaviors are
\begin{equation}
\langle x^2(t) \rangle_{\mathcal{E}} \sim t^3  ~~~~~~~~~~~~~~ \langle [\delta x(t)]^2 \rangle_{\mathcal{E}} \sim t^{7/2},  
\end{equation}
whereas, from Eq.~(\ref{DRIFT0}), we have that
\begin{equation}
 \langle x(t)
\rangle_{\mathcal{E}+\delta\mathcal{E}}-\langle x(t)
\rangle_{\mathcal{E}} \propto \langle x(t) \rangle_{\mathcal{E}} \sim t^{3/2},
\end{equation}
as shown in Fig.~\ref{fig1}. The Einstein relation is therefore
violated in the out-of-equilibrium regime for both the MSD and MSD
around the average current for all the values of the flight time
distribution exponent $g<5$, as summarized in Tab.~\ref{tabella}.

\begin{table*}[!htb]
\begin{center}
\begin{tabular}{|c|c|c|c|}
\hline & $\langle x^2(t) \rangle$ & $\langle x(t) \rangle_{\mathcal{E}}$ & 
 $\langle [\delta x(t)]^2 \rangle_{\mathcal{E}}$ \\ 
\hline $g>5$   & $t$ & $t$ & $t$ \\
\hline $3<g<5$ & $t$ & $t$ & $t^{6-g}$  \\ 
\hline $2<g<3$ & $t^{4-g}$ & $t^{4-g}$ & $t^{6-g}$ \\ 
\hline $1<g<2$ & $t^2$ & $t^2$ & $t^4$ \\ 
\hline
\end{tabular}
\end{center}
\caption{Columns from left to right: MSD for
  unperturbed dynamics; drift in the case of constant field
  $\mathcal{E}>0$; second cumulant, i.e. $\langle [\delta x(t)]^2
  \rangle_{\mathcal{E}} = \langle x^2(t) \rangle_\mathcal{E} - \langle
  x(t) \rangle^2_\mathcal{E}$, in the presence of a constant field
  $\mathcal{E}$.
\label{tabella}}
\end{table*}

\begin{figure}[t!]
\center
\includegraphics[width=.7\columnwidth,clip=true]{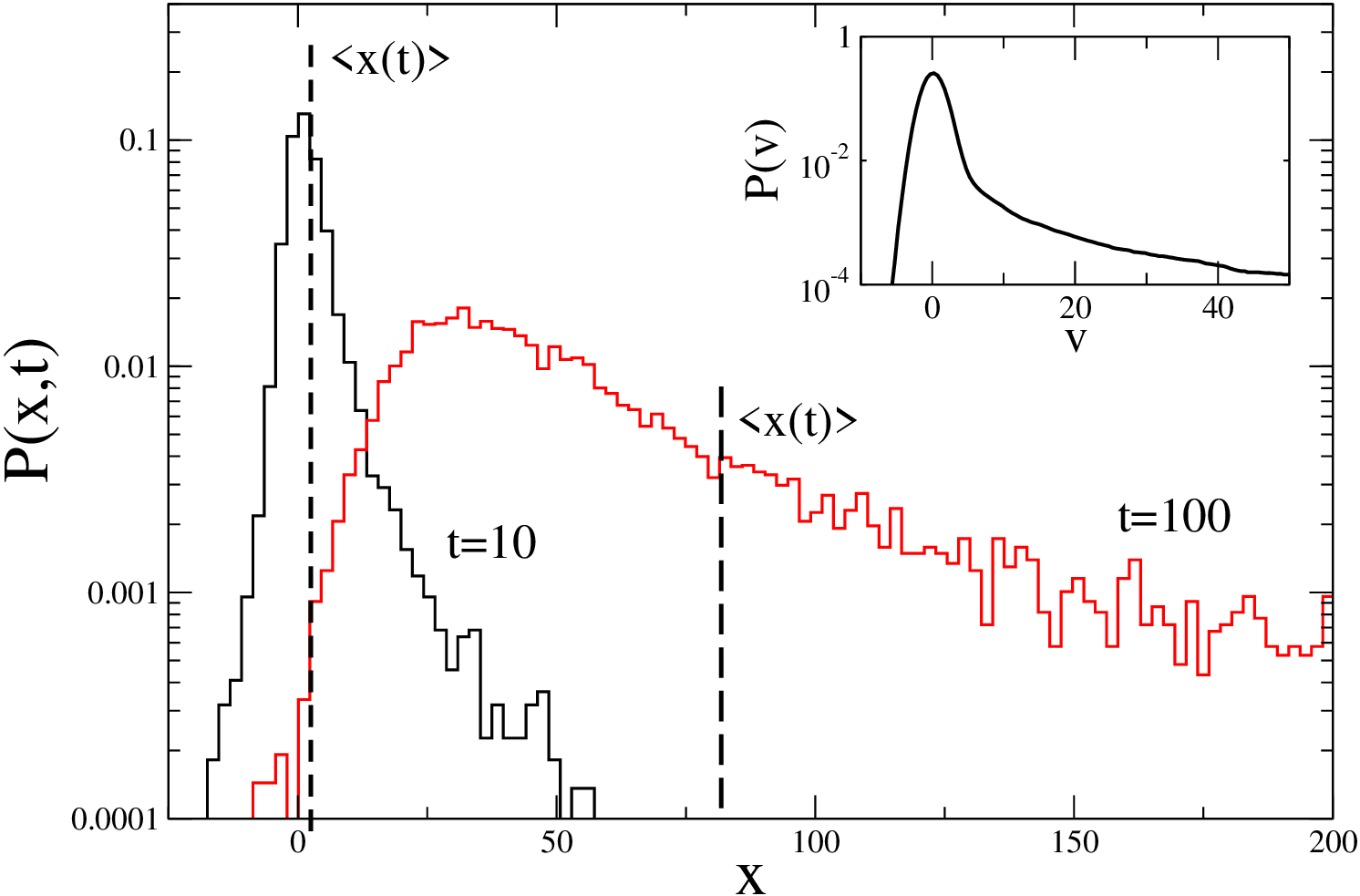}
\caption{Main frame: Semi-Log plot of $P(x,t)$ at $t=10$ (black
  histogram), and $t=100$ (red histogram) for superdiffusive dynamics
  with $g=5/2$ and constant external field $\mathcal{E}=1.5$. Let us
  notice that for increasing time the typical (most frequent) value,
  and the mean (indicated with a vertical dashed line) of the
  distribution become increasingly different for growing times. The
  increasing average value of the distribution is due to the arising
  of a power-law tail for $P(x,t)$. Inset: Semi-Log plot of the
  velocity PDF $P(v)$.}
\label{fig2}
\end{figure}

As already noticed in~\cite{VSGPV11} in the context of subdiffusive
dynamics, such a violation of the Einstein relation is accompanied by
an asymmetric spreading of the PDF $P(x,t)$ of observing the particle
in $x$ at time $t$. In the case of the standard random walk, when a
perturbation $\mathcal{E}$ is applied, the $P(x,t)$ remains Gaussian
and the mean value coincides with the most probable one. On the
contrary, in the present case, the average value of $x(t)$, due to
the strongly asymmetric shape, grows much faster than the most
probable value.  The tail of $P(x,t)$ is reported in
Fig.~\ref{fig2} (main frame).  The stationary distribution of the
velocities is also asymmetric, and with a power-law tail, as shown in
the inset of Fig.~\ref{fig2}. The study of the velocity distribution
in the diffusive case can be found in~\cite{BT12}, while a careful
study of the behavior of higher moments and of the scaling properties
of $P(x,t)$, in similar models in the absence of external field, can
be found in~\cite{ACMV00} and~\cite{BCV10}, respectively.

The study of the Einstein relation on a state with non-zero current
induced by a constant field $\mathcal{E}$ allows us to show that there
is some ``anomaly'' in the dynamics also when the exponent of the
power law distribution of times is $g>3$. More precisely,
when $3<g<5$, at equilibrium, i.e. in the absence of current, a
fluctuation-dissipation experiment would not show any anomaly in the
dynamics, because $\langle [\delta x(t)]^2 \rangle\sim \langle x(t)
\rangle_\mathcal{E}$. On the other hand, the same experiment done out of
equilibrium, i.e. comparing the MSD around the drift with the drift itself, 
shows an evident violation

\begin{equation}
\frac{\langle [\delta x(t)]^2 \rangle_\mathcal{E}}{\langle x(t) \rangle_{\mathcal{E}}}\sim t^{5-g}.
\label{noeinstein}
\end{equation}

\section{Conclusions} 

The present study concerns the comparison of the drift of a probe
particle in the presence of an external field with its MSD, measured
either in the presence or in the absence of the external perturbation.
The outcome of our study is that, while for a L\'evy walk process the
\emph{equilibrium} fluctuation-dissipation relation is always
valid~\cite{BF98}, the situation is very different in the
\emph{out-of-equilibrium} case, where we have a breaking of the
Einstein relation.  In particular, we find that the standard
fluctuation-dissipation relation can be recovered by replacing the MSD
with the diffusion around the average current $\langle x^2(t)
\rangle_\mathcal{E} - \langle x(t) \rangle_\mathcal{E}^2$, as it
happens for standard random walks, \emph{only} for values of the
exponent $g>5$. This is a new an unexpected result, since already for
$g>3$ we observe a simple diffusive dynamics at equilibrium, even if
the $P(x,t)$ is non-Gaussian. The non-Gaussian nature of the
distribution of displacements emerges for $3<g<5$ only through a
``response experiment'' in presence of some currents.

As shown in Fig.~\ref{fig2}, the violation of the
fluctuation-dissipation relation for $g<5$ comes together with a
strongly asymmetric shape of the $P(x,t)$ for large times. Let us note
that such a mechanism, namely a transport mechanism which do not
correspond to a uniform shifts of the $P(x,t)$, is peculiar not only
of this superdiffusive model, but has already been observed in the
context of subdiffusive dynamics~\cite{VSGPV11}.

\ack We thank R. Burioni, A. Vezzani and A. Puglisi for several useful
discussions.  The work of GG and AS is supported by the
``Granular-Chaos'' project, funded by Italian MIUR under the grant
number RBID08Z9JE.




\section*{References}
\bibliographystyle{unsrt}
\bibliography{fluct.bib}

 \end{document}